\documentclass[12pt]{article}
\usepackage{epsf}

\usepackage{a4}

\begin{document}
\newcommand{\ep}{\epsilon}
\newcommand{\fr}{\frac}
\newcommand{\reals}{\mbox{${\rm I\!R }$}}
\newcommand{\nats}{\mbox{${\rm I\!N }$}}
\newcommand{\intgs}{\mbox{${\rm Z\!\!Z }$}}
\newcommand{\cam}{{\cal M}}
\newcommand{\caz}{{\cal Z}}
\newcommand{\cao}{{\cal O}}
\newcommand{\cac}{{\cal C}}
\newcommand{\aaa}{\int\limits_{mR}^{\infty}dk\,\,}
\newcommand{\bbb}{\left[\left(\frac k R\right)^2-m^2\right]^{-s}}
\newcommand{\ccc}{\frac{\partial}{\partial k}}
\newcommand{\fff}{\frac{\partial}{\partial z}}
\newcommand{\iikma}{\aaa \bbb \ccc}
\newcommand{\ddd}{\int\limits_{mR/\nu}^{\infty}dz\,\,}
\newcommand{\eee}{\left[\left(\frac{z\nu} R\right)^2-m^2\right]^{-s}}
\newcommand{\lll}{\frac{(-1)^j}{j!}}
\newcommand{\iinma}{\ddd\eee\fff}
\newcommand{\cah}{{\cal H}}
\newcommand{\nn}{\nonumber}
\renewcommand{\theequation}{\mbox{\arabic{section}.\arabic{equation}}}
\newcommand{\komplex}{\mbox{${\rm I\!\!\!C }$}}
\newcommand{\sip}{\frac{\sin (\pi s)}{\pi}}
\newcommand{\numr}{\left(\frac{\nu}{mR}\right)^2}
\newcommand{\mzs}{m^{-2s}}
\newcommand{\rzs}{R^{2s}}
\newcommand{\abl}{\partial}
\newcommand{\g}{\Gamma\left(}
\newcommand{\ikma}{\int\limits_{\gamma}\frac{dk}{2\pi i}\,\,(k^2+m^2)^{-s}\frac{\partial}{\partial k}}
\newcommand{\ead}{e_{\alpha}(D)}
\newcommand{\sual}{\sum_{\alpha =1}^{D-2}}
\newcommand{\sulnu}{\sum_{l=0}^{\infty}}
\newcommand{\sujnu}{\sum_{j=0}^{\infty}}
\newcommand{\suani}{\sum_{a=0}^i}
\newcommand{\suanzi}{\sum_{a=0}^{2i}}
\newcommand{\zend}{\zeta_D^{\nu}}
\newcommand{\amed}{A_{-1}^{\nu ,D}(s)}
\renewcommand{\and}{A_{0}^{\nu ,D}(s)}
\newcommand{\aid}{A_{i}^{\nu ,D}(s)}
\newcommand{\res}{{\rm res}\,\,\,}
\newcommand{\sn}{\frac{\sin \pi s}{\pi}}
\newcommand{\ha}{\frac{1}{2}}
\newcommand{\smj}{\sum_{j=\ha}^\infty (2j+1)}
\newcommand{\smN}{\sum_{k=1}^N}
\newcommand{\smk}{\sum_{k=0}^\infty \frac{(-1)^k}{k!}}
\newcommand{\sma}{\sum_{a=0}^{2i} x_{i,a}}
\newcommand{\itz}{\int_{\frac{mR}j}^\infty dz}
\newcommand{\paran}{\left[\left(\frac{zj}R \right)^2-m^2\right]}
\newcommand{\dz}{\frac{\partial }{\partial z}}
\newcommand{\bee}{\begin{equation}}
\newcommand{\bea}{\begin{eqnarray}}
\newcommand{\eea}{\end{eqnarray}}
\newcommand{\app}[1]{\section{#1}\renewcommand{\theequation}
        {\mbox{\Alph{section}.\arabic{equation}}}\setcounter{equation}{0}}
\def\beq{\begin{eqnarray}}
\def\eeq{\end{eqnarray}}

\begin{titlepage}

\title{\begin{flushright}
\end{flushright}
\vspace{3mm}
{\Large \bf Casimir energy in the MIT bag model}}
\author{
E. Elizalde\thanks{E-mail address:
eli@zeta.ecm.ub.es, elizalde@io.ieec.fcr.es}\\
Unitat de Recerca, CSIC, IEEC, Edifici Nexus 104,\\
Gran Capit\`a 2-4, 08034 Barcelona, Spain\\ and
Departament ECM and IFAE, Facultat de F\'{\i}sica, \\
Universitat de Barcelona, Diagonal 647, 08028 Barcelona, Spain \\
M. Bordag\thanks{E-mail address: michael.bordag@itp.uni-leipzig.de} and 
K. Kirsten\thanks{E-mail address: klaus.kirsten@itp.uni-leipzig.de}\\
Universit{\"a}t Leipzig, Institut f{\"u}r Theoretische Physik,\\
Augustusplatz 10, 04109 Leipzig, Germany
}

\thispagestyle{empty}

\vspace*{-1mm}

\maketitle

\vspace*{-2mm}

\begin{abstract}
The vacuum  energies corresponding to massive Dirac
fields with the boundary conditions of the MIT bag model
are obtained. 
The calculations are done with the fields occupying the regions
inside and outside the bag, separately. The renormalization procedure for 
each of the situations is studied in detail, in particular
the differences occurring with respect to the case when the field extends over
the whole space. The final result contains several constants undergoing
 renormalization, which can be determined only experimentally.
The non-trivial finite parts which appear in the massive case are found
exactly, providing a precise determination of the complete, renormalized
zero-point energy for the first time, in the fermionic case. The vacuum 
energy behaves like inverse powers of the mass for large masses.
\end{abstract}

\vfill \noindent PACS: 11.10.Gh, 02.30.-f \\
Running title: Casimir energy in the MIT bag

\end{titlepage}
\section{I. Introduction}
\setcounter{equation}{0}
The first modern calculation of the vacuum energy density of a quantum
field in the presence of boundaries is almost fifty years old.
As is well known, it is due to H. Casimir \cite{casimir48}.
Its first measurable consequence was the
attraction in the electromagnetic
vacuum of two neutral, infinitely conducting plates (thereafter called
Casimir effect, see for instance Ref. \cite{greiner}).
There had been, before that time, other calculations and explanations
of the attraction of two neutral bodies, understanding them as mere 
van der Waals effects, \cite{vdW1},  and
also a paper by Casimir and Polder \cite{cp1} where the finiteness of the
 velocity of light had been  taken into account.  Casimir's paper
\cite{casimir48} was the first one where an absolutely modern
quantum field theoretical
calculation was performed,  using the concept 
 of zero-point energy (whose physical relevance
was somehow unclear at that time). The treatment of the divergences 
resulting from the infinitely many degrees of freedom has been the most
difficult point. Since then,
calculations of the vacuum energy 
have attracted the interest of many scientists because
it turns out that, in
different  contexts,
 the inclusion of quantum fluctuations about semi-classical
configurations is essential.
On the other hand,
spherically symmetrical situations are very important for practical
applications. The calculations involved are certainly much
more complicated than in the case of systems with plane boundaries.

Historically the first far reaching
ideas involving vacuum energies in the case of spherical configurations
also originated with Casimir.
He proposed that the force
stabilizing a classical electron model arises from the zero-point
energy of the electromagnetic field within and without a perfectly
conducting spherical shell \cite{casimir56}. Having found an
attractive force between parallel plates due to the vacuum energy
\cite{casimir48}, the hope was that the same would occur for the
spherically symmetric situation. Unfortunately, as Boyer 
\cite{boyer68} first showed, for this geometry the stress is
repulsive \cite{baldup78,mil78}. Nowadays it is known that the 
Casimir energy depends strongly on the geometry of the space-time
and on the boundary conditions imposed. This is  a very active field
of research (see, for instance, \cite{eorbz},\cite{ee}).

More recently the zero-point energy has received considerable attention 
in the context
of the bag model \cite{cho74}-\cite{francia} and chiral bag model
\cite{vep90}-\cite{hor3}.
In these systems, quarks and gluons are free inside the bag, but are
absolutely confined to it, being unable to cross the boundary surface.
This is imposed, mathematically, by appropriate boundary conditions. 
The sum of the mesonic, valence quark and vacuum quark contributions to
the baryonic number have been found to be independent of the bag radius
and pion field strength, being the vacuum quark contributions
---which are analogues of the Casimir effect in QED--- essential in
the calculation of baryonic observables. The issue of regularization
in this model is certainly non-trivial. Under specific circumstances,
different regularization procedures can yield different results and
real physical problems arise in the calculation of quark vaccum
 contributions to some barionic observables, as the energy itself.

The Casimir energy for a spherical capacitor is usually given by the sum
of two terms: an internal and an external one.
Volume divergences cancel
in each of the two regions separately, but surface and curvature
divergences survive in each part and only cancel when interior and 
exterior contributions are added up, leaving then constant terms.
In chiral bag models an additional divergence --logarithmic in the
 plasma frequency--- appears. Having done these considerations, one should
observe, however, that in the bag model quarks are supposed to exist
in the {\it interior} of the bag only and, therefore, there is no clear way
of eliminating these surface, curvature and logarithmic divergences.
Some authors (see \cite{vep90}) have suggested a revision of the
Cheshire cat principles in the sense of including free quarks as
correct high-energy degrees of freedom for the bag exterior, which would
appear with a smooth transition at energies of the order of some GeV.
In this way divergences would cancel quite naturally.
However, in our opinion this deconfining transition is even a 
harder issue
than those of regularization and renormalization. It will not be dealt
 with here.

For massless fermions
the zero-point energy was considered some time ago in 
Refs. \cite{milton80,milton83}, finite temperature effects were
taken into account in \cite{francia,hor2}.
The massless fermionic field inside and outside the spherical bag
was analysed  in \cite{milton83}. In the last case, a mutual 
cancellation of the
divergences of the inner and outer spaces occurs. As a result, finite 
zero-point energies are found. However, when considering only the 
inner space, divergences
arise  and it is necessary to introduce contact terms and perform a
renormalization of their coupling. Results for the massive fermionic 
fields contain new ultraviolet-divergent terms in addition to those 
occurring in the massless case, as has been discussed
in \cite{baacke83}. Further considerations, especially on the
 renormalization procedure 
---necessary in order to carry out
these calculations--- and also on its precise physical
 interpretation, can be found in \cite{andreas}.

In most of the  papers mentioned above a Green's function approach has been 
used in order to calculate the zero-point energy. An exception is Ref.
\cite{andreas}, where, in the general setting of an ultrastatic spacetime
with or without boundaries, a systematic procedure which makes use of
 zeta function
regularization was developed. In this approach, a knowledge of the zeta
function of the operator associated with the field equation together
with (eventually) some appropriate
 boundary conditions is needed. Recently, a detailed
description of how to obtain the zeta function for a massive scalar field
inside a ball satisfying Dirichlet or Robin boundary conditions has been
given elsewhere by  the
authors of the present work \cite{bk,bek}. An analytical continuation
to the whole complex plane was obtained there and subsequently
 applied to the computation of an
arbitrary number of heat-kernel coefficients. In  ensuing papers
 \cite{begk,bdk} the functional
determinant was considered too
 and, furthermore, the method
has been also applied to spinors \cite{abdk} and p-forms 
\cite{eli1,eli2,stuart}.
All the above considerations are purely analytical and quite 
precise. In order to 
obtain  explicit values for the Casimir energy, however,
 a numerical evaluation of an integral was
necessary. This has been  done in different cases, in particular
 for the massless scalar field and the
electromagnetic field \cite{rom}, partly reobtaining previous
results.

To finish this description of recent previous work, let us mention that in
Ref. \cite{bekl96} we have investigated the case of a massive scalar field
in the bag. We have discussed there how, for the case of a massive field
---already for a {\it scalar} one---
 non-trivial finite parts which depend on an adimensional variable
involving the mass are present, that need to be properly renormalized, in 
order to get the corresponding zero-point energy. In the present paper we 
shall extend our analysis to the case of Dirac fields, generalizing in this 
way our considerations to a situation that approaches very much the conditions
of a realistic MIT bag model. 

The organization of the paper is as follows. 
We shall rely on our previous work (dealing with the bosonic case)
for 
a precise description of our method ---which was given there in full detail
\cite{bekl96}--- as well as for the particular  formulas that are needed 
in the subsequent study of
 the zeta function of the problem we consider here. 
We feel that to repeat all this here would not be justified. Consequently, 
in Sect. 2 we will proceed already with the specific description of the model 
for the case 
of Dirac fields  inside the bag with boundary conditions 
corresponding to the MIT bag. 
Starting from the Dirac equation and imposing the boundary conditions we 
will derive an eigenvalue equation in terms of Bessel functions. This will 
be the basic equation to solve, what we shall do in the same section for the 
interior of the bag. In section 3 we will explain the renormalization
scheme used for the model. Sect. 4 contains the analogous treatment
for the region exterior to the bag and for the whole space.  
Adding
up the interior and the exterior contributions, we will see 
how the divergences cancel among themselves, as well as the
influence of this cancellation on the
compulsory renormalization process. 
It turns out that 
important differences with respect to the non-fermionic case appear
concerning this issue, although we shall argue that, in the end, they will 
not affect substantially the interpretation of the physical results. 
Sect. 5 is devoted to conclusions. 
The appendixs contain some hints and technical details that have been
 used in the  derivation of the zeta function (App. A) 
and a full list of the 
constituents that build up the subtraction terms in the decomposition of
the zeta function, an essential (although rather technical) step
in our method (App. B).

\section{Fermions inside the bag}
\setcounter{equation}{0}
The first task is to derive the energy eigenvalue equations for a 
Dirac spinor subject to the MIT bag boundary conditions. The setting we 
consider first is the Dirac spinor inside a spherically symmetric bag
confined to it by the appropriate boundary conditions. 
The coordinates we  use are
just the spherical ones, $r,\theta,\varphi$, which best adapt to the form
of the bag. Thus, we must solve the equation:
\beq
H\phi_n (\vec r) = E_n \phi_n (\vec r) , \label{2.1}
\eeq
$H$ being the Hamiltonian,
\beq
H=-i\gamma^0 \gamma^{\alpha} \partial _{\alpha} +\gamma^0 m, \label{2.2}
\eeq
with       the boundary conditions
\beq
\left[1+i \left(\frac{\vec r } r \vec \gamma \right) \right] \phi_n 
\left|_{r=R} \right. =0 .\label{2.3}
\eeq
These boundary conditions guarantee that no quark current is lost through
the boundary.

The  separation to be carried out
 in the eigenvalue equation (\ref{2.1}) is rather
standard and will not be given here in detail. Let $\vec J$ be the 
total angular momentum operator and $K$ the spin projection operator. 
Then there exists a simultaneous set of eigenvectors of $H,\vec J^2,J_3,K$
and the parity $P$. The eigenfunctions for positive eigenvalues $\kappa
=j+1/2$ of $K$ read
\beq
\phi_{jm} = \frac A {\sqrt{r} } \left(
   \begin{array}{l} i J_{j+1} (\omega r ) \Omega_{jlm} \left( \frac {\vec r} r 
\right)  \\
    - \sqrt{\frac{E-m}{E+m} } J_j (\omega r ) \Omega_{jl'm } 
         \left( \frac {\vec r} r \right)
   \end{array} 
\right), \label{2.4}
\eeq
whereas, for $\kappa = -(j+1/2) $, one finds
\beq
\phi_{jm} = \frac A {\sqrt{r} } \left(
   \begin{array}{l} i J_{j} (\omega r ) \Omega_{jlm} \left( \frac {\vec r} r
\right)  \\
     \sqrt{\frac{E-m}{E+m} } J_{j+1} (\omega r ) \Omega_{jl'm }
         \left( \frac {\vec r} r \right)
   \end{array}
\right). \label{2.5}
\eeq
Here 
$\omega=\sqrt{E^2-m^2}$, $A$ is a normalization constant and 
$\Omega_{jlm} (\vec r /r)$ are the well known spinor harmonics. 
In order to obtain eigenfunctions of the parity operator we must set
 $l'=l-1$ in
(\ref{2.4}) and $l'=l+1$ in (\ref{2.5}). In both cases, $j=1/2,3/2,...,
\infty$, and the eigenvalues are degenerate in $m=-j,...,+j$.

Imposing the boundary conditions (\ref{2.3}) on the solutions 
(\ref{2.4}) and (\ref{2.5}), respectively, one easily finds the 
corresponding implicit eigenvalue
equation. For $\kappa > 0$, it reads
\beq
\sqrt{\frac{E+m}{E-m}} J_{j+1} (\omega R) +J_j (\omega R) =0 , \label{2.6}
\eeq
and for $\kappa < 0$, on its turn, 
\beq
J_j (\omega R)- \sqrt{\frac{E-m}{E+m}} J_{j+1} (\omega R) =0 .\label{2.7}
\eeq
Regretfully, it is not possible to find an explicit solution of
equations (\ref{2.6}) 
and (\ref{2.7}). But as we have shown in our previous paper for the case 
of the scalar field ---and  will describe below for the spinor field---
 the information 
displayed in (\ref{2.6}) and (\ref{2.7}) is already enough for the 
calculation of the ground state energy for massive spinors in the bag.

The regularization of this ground state energy will be performed by
 using the zeta function method. In short, we consider
\beq
E_0 (s) &=& - \frac{1}{2} \sum_k 
\left( E_k^2 \right)^{1/2 -s} \mu^{2s}, \qquad
\mbox{Re}\ s >s_0= 2\nonumber\\
&=& -\frac 1 2 \zeta^{(int)}(s-1/2) \mu^{2s} \label{grounden},
\eeq
and later analytically continue to the value $s=0$ in the complex plane.
Here $s_0$ is the abscissa of convergence of the series, $\mu$ the usual 
mass parameter and 
\beq
\zeta^{(int)} (s) = \sum_k (E_k^2) ^{-s} \label{relzeta}
. 
\eeq
The power of the method lies in the well defined prescriptions 
and procedures that we have at our hand to analytically continue the series
to the rest of the complex $s$-plane, even when the spectrum $E_k$ is
not known explicitly (as will in fact be the case).
These procedures have been developed and described in great detail in
\cite{bk,bek,bekl96}, so that we can be brief here.

The zeta function in the interior space is given by
\beq
\zeta^{(int)} (s) &=& 2 \sum_{j=1/2,3/2,\ldots}^\infty (2j+1) \int_\gamma \frac{dk}{2\pi i} 
(k^2 +m^2)^{-s} \nn \\ && \hspace{-10mm} 
\times \, \frac{\partial}{\partial k}
 \ln \left[ J_j^2 (kR) - J_{j+1}^2 (kR) 
+ \frac{2m}{k} J_j (kR) J_{j+1} (kR) \right].
\eeq
Here the factor of $2$ results from taking into account particles and 
antiparticles.
Using the method ---ordinarily employed in this situation--- of deforming the 
contour which originally encloses the singular points on the real axis,
until it covers the imaginary axis, 
after simple manipulations we obtain the following equivalent expression for 
$\zeta^{(int)}$:
\beq
\zeta^{(int)}   (s) &=& \frac{2 \sin \pi s}{\pi} 
\sum_{j=1/2,3/2,\ldots}^\infty (2j+1) 
\int_{mR/j}^\infty dz \, \left[ \left( \frac{zj}{R}\right)^2 -m^2
\right]^{-s}  \\
&& \times \frac{\partial}{\partial z} \ln\left\{z^{-2j} \left[ I_j^2 (zj)
 \left( 1 + \frac{1}{z^2}- \frac{2mR}{z^2j} \right) + {I_j'}^2 (zj)
\right. \right. \nn \\ && \left. \left.
 + \frac{2R}{zj} \left( m - \frac{j}{R} \right) I_j (zj) {I_j}' (zj)
\right]\right\}. \nn
\eeq

As is usual, we will now split the zeta function into two parts:
\beq
\zeta^{(int)} (s)=Z_N(s)+\sum_{i=-1}^N A_i(s),
\eeq 
namely a regular one, $Z_N$, and a remainder that contains the 
contributions of the $N$ first terms
of the Bessel functions $I_\nu (k)$ as $\nu, k \rightarrow \infty$ with 
$\nu /k$ fixed 
\cite{abramo}. The number $N$ of terms that have to be subtracted is in
general the minimal one necessary in order to absorb all possible divergent
contributions into the groundstate energy, Eq.~(\ref{grounden}).
In our case, $N=3$. This is a general procedure, commonly applied in order to
deal with such kind of divergences. We get
\beq
Z_3(s)&=& 2\sn \smj \itz \paran^{-s}  \nn\\
&& \hspace{-17mm}\times \frac{\partial}{\partial z}
 \left\{ \ln \left[I_j^2(zj)(1+\frac{1}{z^2}-\frac{2mR}{z^2j})+
I'_j\!{}^2(zj)+\frac{2R}{zj} (m-\frac{j}{R}) I_j(zj)I'_j(zj)\right] \right. 
 \nn\\
&&-\left.\ln\left[\frac{e^{2j\eta}(1+z^2)^\ha(1-t)}{\pi j z^2}\right]
   -\sum_{k=1}^3 \frac{D_k(mR,t)}{j^k}\right\},
\eeq
where $\eta = \sqrt{1+z^2} +\ln [z/(1+\sqrt{1+z^2})]$ and $t=1/\sqrt{1+z^2}$.
After renaming $mR=x$, the relevant polynomials are given by
\beq
D_1(t)&=&{\frac {{t}^{3}}{12}}+\left (x-1/4\right )t \nn\\
D_2(t)&=&-{\frac {{t}^{6}}{8}}-{\frac {{t}^{5}}{8}}+\left (-{\frac {x}{2}}+1/8
\right ){t}^{4}+\left (-{\frac {x}{2}}+1/8\right ){t}^{3}-{\frac {{t}^
{2}{x}^{2}}{2}} \nn\\
D_3(t)&=&{\frac {179\,{t}^{9}}{576}}+{\frac {3\,{t}^{8}}{8}}+\left (-{\frac {23
}{64}}+{\frac {7\,x}{8}}\right ){t}^{7}+\left (x-1/2\right ){t}^{6}+
\left ({\frac {9}{320}}-{\frac {x}{4}}+{\frac {{x}^{2}}{2}}\right ){t}
^{5}    \nn\\
&&+\left ({\frac {{x}^{2}}{2}}+1/8-{\frac {x}{2}}\right ){t}^{4}+
\left (-{\frac {x}{8}}+{\frac {5}{192}}+{\frac {{x}^{3}}{3}}\right ){t
}^{3}. \label{asympol}
\eeq
The asymptotic contributions $A_i (s)$, $i=-1,...,3$, are defined as
\beq
A_{-1} (s) &=& 
\frac{8\sin (\pi s)} {\pi} \sum_{j=1/2}^{\infty} 
j(j+1/2) \int_{mR/j} ^{\infty} \left( \left( \frac{xj} R \right)^2 -m^2 
\right)^{-s}
\frac{\sqrt{1+x^2} -1} x \nonumber\\
A_0 (s) &=& 
\frac{4\sin (\pi s)} {\pi} \sum_{j=1/2}^{\infty}
(j+1/2) \int_{mR/j} ^{\infty} \left( \left( \frac{xj} R \right)^2 -m^2 
\right)^{-s}
\frac{\partial} {\partial x} \ln \frac{\sqrt{1+x^2} (1-t) } {x^2} 
\nn\\
A_i (s) &=& 
\frac{4\sin (\pi s)} {\pi} \sum_{j=1/2}^{\infty}
(j+1/2) \int_{mR/j} ^{\infty} \left( \left( \frac{xj} R \right)^2 -m^2 
\right)^{-s}
\frac{\partial} {\partial x}
\frac{D_i (t)} {j^i}.  \label{aisanf}
\eeq
The small mass expansion can be conveniently represented as
\beq
A_{-1}(s)& =&\frac{R^{2s}}{\sqrt{\pi} \,\Gamma (s)}
 \sum_{k=0}^\infty \frac{(-1)^k}{k!} (mR)^{2k} \frac{\Gamma( k+s-\ha)}{k+s}
 \nn \\ && \times
\left[2\zeta(2k+2s-2,\ha)+\zeta(2k+2s-1,\ha)\right] 
\nn \\
A_{0}(s)&=&-\frac{R^{2s}}{2\sqrt{\pi} \, \Gamma (s) } \sum_{k=0}^\infty 
\frac{(-1)^k}{k!}  (mR)^{2k}
	\frac{\Gamma ( k+s+\ha)}{k+s}  \nn \\ && \times
	\left( 2\zeta(2k+2s-1,\ha)+\zeta(2k+2s,\ha) \right) 
\nn \\
A_{i}(s)&=&-\frac{2R^{2s}}{\Gamma (s)} \sum_{k=0}^\infty
 \frac{(-1)^k}{k!} (mR)^{2k}
	\left[ 2\zeta(2k+2s+i-1,\ha)+\zeta(2k+2s+i,\ha) \right]  \nn\\
&&\times \sum_{a=0}^{2i} x_{i,a} \frac{ \Gamma (k+s+\frac{a+i}{2})}{
\Gamma ( \frac{a+i}{2})} .
\eeq
In this expression, the $x_{i,a}$ are the coefficients of the expansion 
of the functions $D_i(t)$, i.e.,
\beq
D_i(t)=\sma t^{a+i}.
\eeq
Note that here we encounter the same problem that occurred already in
the scalar case.  One needs  a representation that is useful and valid for 
an (in principle) arbitrary value of $m$. To this end one can actually
 proceed in different ways, casting the final result in terms of 
convergent series or integrals. Our {\it leitmotiv} 
will be the following: 
we will always try to express the final result in terms of the formula which 
is more appropriate for practical evaluation (e.g., numerical, in general).
 This means that, sometimes,  instead of having the closed convergent sums 
that could be universally used in the scalar case,  rapidly 
converging integrals better suited for numerical 
analysis will be here preferred. 

With this aim, we note that
after performing the z-integration 
the $A_i (s)$, for $i\geq 1$, can be written in the following form,
\beq
A_i (s) &=& -\frac{4m^{-2s}}{\Gamma (s)} \sum_{a=0}^{2i} \frac{x_{i,a}}
{(mR)^{i+a} } \frac{\Gamma (s+(i+a)/2) }{\Gamma ((i+a)/2 ) } \nn\\
& &\times \left[ f (s;1+a; (i+a)/2 ) +\frac 1 2 f (s;a; (i+a)/2 ) \right],
\label{ais} 
\eeq
with the definition 
\beq 
f(s;a;b) =\sum_{\nu = 1/2,3/2,\ldots} ^{\infty} 
\nu ^a \left( 1+ \left( \frac{\nu} 
{mR } \right ) ^2 \right) ^{-s-b}. \label{fab} 
\eeq
The remaining thing to do in the present case is to calculate the
$f(s;a;b)$ for the relevant values at $s=-1/2$. 
This is a systematic calculation that will be sketched
in App. A.
Let us here mention only that an essential step is to use the simple
recurrence:
\beq
f(s;a;b) = (mR)^2 \left[ f(s;a-2;b-1)- f(s;a-2;b)\right].
\label{recur}
\eeq
In App. B we give the whole list of starting terms that, in
addition to the recurrence formula (\ref{recur}), are
strictly necessary for obtaining explicitly all the $A_i(s)$
needed in our calculation.

\section{Discussion of the renormalization}
\setcounter{equation}{0}
For the discussion of the renormalization let us look for the divergent
terms in the groundstate energy. By construction they are all contained 
in the contributions $A_i (s)$. Having their explicit form at hand 
(see Apps.      A and B) they can be given quickly. In particular, we have,
 for the interior part:
\beq
\res A_{-1}^{(int)} (-1/2) &=& -\frac{m^4R^3}{6\pi} +\frac{m^2 R} {12\pi} 
+\frac 7 {480\pi R} \nonumber\\
\res A_0^{(int)} (-1/2) &=& -\frac{m^2 R} {2\pi} -\frac 1 {24 \pi R} \nn\\
\res A_{1}^{(int)}(-1/2) &=& -\frac{m^3 R^2} {\pi} +\frac{m^2 R}{ 12\pi} 
+\frac m {12\pi} -\frac 1 {48\pi R} \nn\\
\res A_2^{(int)} (-1/2) &=& -\frac{m^2 R} 4 -m \left( \frac 1 {8} +\frac 1 
{2\pi} \right) +\frac 1 {128 R} +\frac 1 {24 \pi R} \nn\\
\res A_3^{(int)} (-1/2) &=& \frac{2m^3 R^2}{ 3\pi} +m^2 R \left (
 \frac 1 4 +\frac 
2 {3\pi} \right) \nn\\ 
& &+m \left( \frac 1 {8 } +\frac 7 {20 \pi} \right) 
-\frac 1 {128 R} -\frac{97} {10080 \pi R} \nn
\eeq
and, as a result, 
\beq
\mbox{Res}\ \zeta^{(int)} (-1/2) = -\frac{1}{63  \pi R} -\frac{m}{15 \pi}
+\frac{m^2R}{3 \pi}  -\frac{m^3R^2}{3\pi}  -\frac{m^4R^3}{6 \pi}.
\label{poles}
\eeq
These terms form the minimal set of counterterms necessary in order to
renormalize our theory.

In the scalar case one had the peculiar situation
that there were
no divergent contributions of the form $\sim m^3,m$ in the zeta function
description \cite{bekl96} ---although
in other regularizations they indeed appear 
\cite{andreas}. So in principle one had the choice of renormalizing the 
associated couplings. Contrarily, for spinors, 
as seen in (\ref{poles}), the
coupling constants of {\it all} terms appearing have to be renormalized. 
We are led into a physical system consisting of two parts:
\begin{enumerate}
\item A classical system consisting of a spherical surface ('bag') with radius
 $R$. Its energy reads:
\beq
E_{class} = p V + \sigma  S + F R +k +\frac{h}{R} ,\label{n1}
\eeq
where $V=\frac{4}{3}\pi R^3$ and $S=4\pi R^2$ are the volume and the surface
of the bag,
 respectively. This energy is determined by the parameters: $p$  pressure,
$\sigma$  surface tension, and $F$, $k$, and $h$ which do not have special
 names. 
\item A spinor quantum field $\varphi (x)$ obeying the Dirac 
equation 
and the MIT boundary conditions 
(\ref{2.3}) on the surface.  
The quantum field has a ground state energy given by $E_0$, Eq.
(\ref{grounden}).    
\end{enumerate}
Thus, the complete energy of the physical system is
\beq
E= E_{class} +E_0 \label{eges}
\eeq
and in this context the renormalization can be achieved by shifting the 
parameters in $E_{class}$ by an amount which
cancels the divergent contributions. 

First we perform a kind of minimal subtraction, where only the 
divergent contribution is eliminated,
\beq
p &\to &p - \frac{m^4} {16\pi^2}
\frac 1 {s}
\quad
\sigma  \,\to  \, \sigma -\frac{m^3}{24\pi^2} \frac 1 s\nn\\
F  &\to &  F + \frac{m^2}{6\pi} 
 \frac 1 {s}
\quad
k\, \to \, k-\frac m {30\pi } \frac 1 s
\label{n8}\\
h& \to & h- \frac 1 {126\pi }  
 \frac 1 {s}.\nn
\eeq
The quantities $\alpha = \{p,\sigma, F,k,h\}$ are a set of free parameters of
the theory to be determined experimentally. In principle we are free
to perform finite renormalizations at our choice of all the parameters. 

We find natural to perform two further renormalizations.
First it is possible
to determine the asymptotic behavior of the $A_i$ for $m\to\infty$ using the 
results of Apps.      A and B. The finite pieces not vanishing in the
limit $m\to \infty$ are all of the same type appearing in the classical 
energy. Our first finite renormalization is such that those pieces are 
cancelled. As a result, only the "quantum contributions" are finally included, 
because, physically, a quantum field of infinite mass is not expected 
to fluctuate. The resulting $A_i$ will be  called $A_i ^{(ren)}$. 

Concerning $Z_3$ we have not been able to determine analytically 
its complete 
non-vanishing behavior for $m\to\infty$. Instead, 
for the numerical analysis, as shown in Fig. 1, we
have constructed a numerical fit of $Z_3$ by a polynomial 
of the form
\beq
P(m) = \sum_{i=0}^4 c_i m^i, \nn
\eeq
and then subtracted this polynomial from $Z_3$. As explained above, 
this is nothing
else than an ulterior finite renormalization. The result will be 
denoted by $Z_3^{(ren)}$.

Summing up, we can write the complete energy as
\beq
E=E_{class} +E_0^{(ren)}, \label{egesend}
\eeq
where $E_{class}$ is defined as in  
(\ref{n1}) with the renormalized parameters $\alpha$ 
and $E_0^{(ren)}= Z_3^{(ren)} + \sum_{i=-1}^3 A_i^{(ren)}$.

\begin{figure}[h]
\epsffile{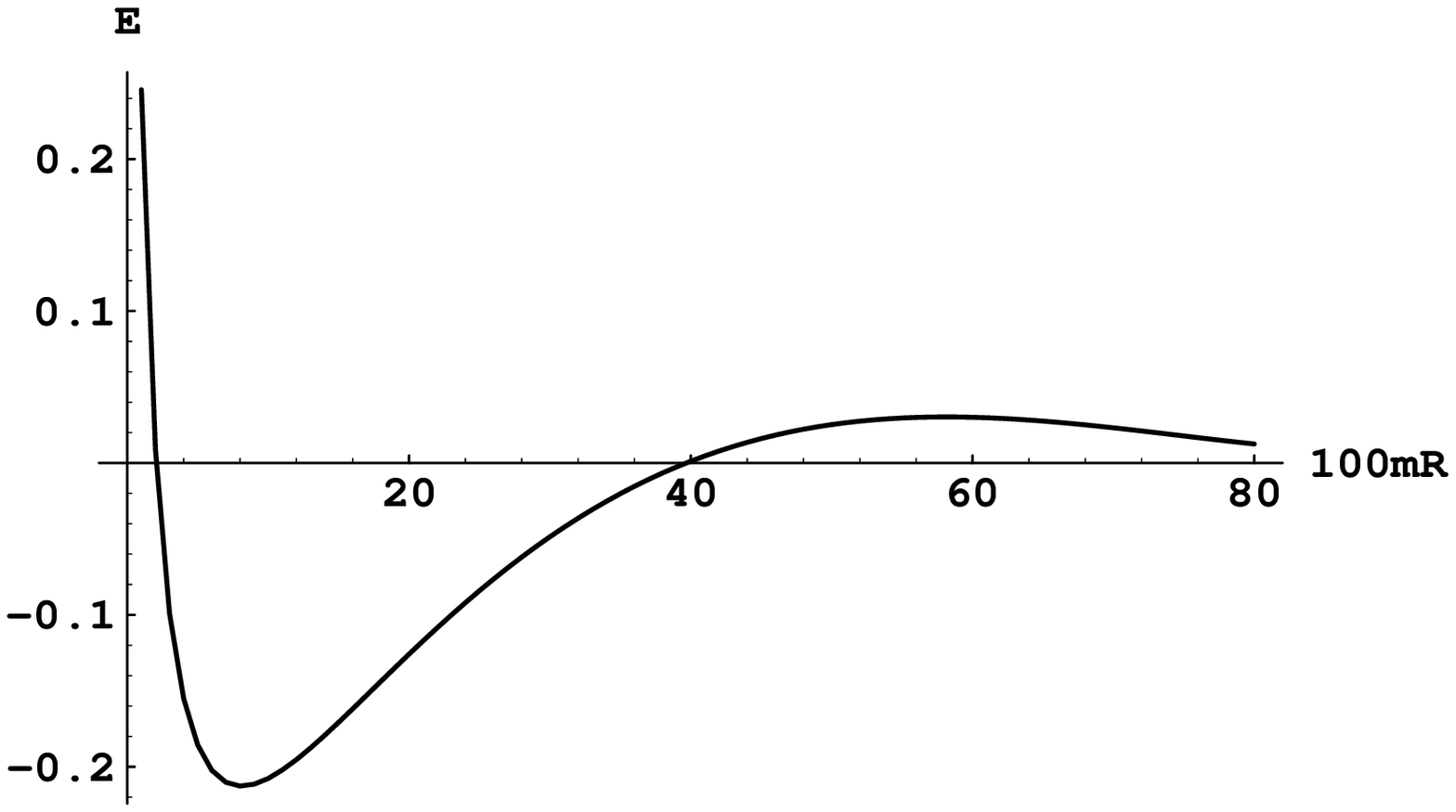}
\caption{ The energy  $E_0^{(ren)}$  as a function of the radius for a specific choice of parameters.  }
\end{figure}

Figure 1 shows the numerical analysis of the energy 
$E_0^{(ren)}$ of the system for this
specific choice of renormalization. The energy exhibits a clear minimum
corresponding to a stability bag radius.

\section{Exterior of the bag and a model for the whole space}
\setcounter{equation}{0}
The analysis of the region exterior to the bag is quite similar to the
one carried out for the interior region. Only some specific differences
appear both in the formulas and in the results. The expression for the zeta
function in the exterior region is essentially
the same as the one corresponding to
the interior, but for the replacement of the Bessel $I_j$ functions with
 Bessel
$K_j$ functions, namely
\beq
\zeta^{(ext)} (s) &=& \frac{2 \sin \pi s}{\pi}
 \sum_{j =1/2,3/2,\ldots}^\infty (2j+1) \int_{mR/j}^\infty dz \,
\left( (zj/R)^2 -m^2 \right)^{-s}
 \nn \\ && \times \, \frac{\partial}{\partial z}
 \ln \left[ z^{2j} \left( K_j^2 (zj) - K_{j+1}^2 (zj)
+ \frac{2mR}{zj} K_j (zj) K_{j+1} (zj) \right) \right]\nn.
\eeq
The splitting of the zeta function has also the same aspect as for the
interior region. We have, in particular
\beq
A_{-1}^{(ext)} (s) = A_{-1}^{(int)} (s),  \nn
\eeq
\beq
A_0^{(ext)} (s) &=& \frac{4\sin (\pi s)} {\pi} 
\sum_{j=1/2}^{\infty} (j+1/2) \int_{mR/j}^\infty dz \,
\left( (zj/R)^2 -m^2 \right)^{-s}
\nn \\ && \, \qquad \qquad \times \frac{\partial}{\partial z}
 \ln \left[\frac{1+t} t\right], \nn
\eeq
and the polynomials that replace the $D_i(t)$ above are here
($x=mR$)
 \beq
\overline{D}_1(t)& =& {t\over 4} +  x\,t - {{{t^3}}\over {12}}, \nn \\
 \overline{D}_2(t) & =& {{- x^2 \,{t^2}  }\over 2} -
   {{{t^3}}\over 8} - {{ x \,{t^3}}\over 2} + {{{t^4}}\over 8} +
   {{ x\,{t^4}}\over 2} + {{{t^5}}\over 8} - {{{t^6}}\over 8}, \nn \\
 \overline{D}_3(t)& =& {{-5\,{t^3}}\over {192}} - {{x \,{t^3}}\over 8} +
   {{{{x }^3}\,{t^3}}\over 3} + {{{t^4}}\over 8} +
   {{x \,{t^4}}\over 2} + {{{{x }^2}\,{t^4}}\over 2} -
   {{9\,{t^5}}\over {320}} \nn \\ && - {{x \,{t^5}}\over 4} -
   {{{{x }^2}\,{t^5}}\over 2} - {{{t^6}}\over 2} - x \,{t^6} +
   {{23\,{t^7}}\over {64}} + {{7\,x \,{t^7}}\over 8} +
   {{3\,{t^8}}\over 8} - {{179\,{t^9}}\over {576}} \nn.
\eeq
As for the functions $A_i^{(ext)} (s)$, one obtains the same expressions
as before, with the  replacement  of the polynomials
$D_i(t)$ with the corresponding polynomials $\overline{D}_i(t)$.

In principle,  the same procedure as before
 can be applied now in order to get 
an analytical expression for the whole energy of the exterior space. 
Instead, we want to restrict ourselves here to the specific changes 
that show up
when discussing the renormalization. For doing that, we only have to
consider the pole of the different $A_i^{(ext)}$. In particular, we have
for the residua
\beq
\mbox{res}\ A_{-1}^{(ext)} (-1/2) &=& -\frac{m^4R^3}{6\pi}+\frac{m^2 R} {12\pi}
+\frac 7 {480\pi R} = \mbox{res}\ A_{-1}^{(int)} (-1/2)  \nonumber\\
\mbox{res}\ A_0^{(ext)} (-1/2) &=& \frac{m^2 R} {2\pi} +\frac 1 {24 \pi R}
= -\mbox{res}\ A_0^{(int)} (-1/2) \nn\\
\mbox{res}\ A_{1}^{(ext)}(-1/2) &=& -\frac{m^3 R^2} {\pi} -\frac{m^2 R}{ 12\pi}
+\frac m {12\pi} +\frac 1 {48\pi R} \nn\\
\mbox{res}\ A_2^{(ext)} (-1/2) &=& -\frac{m^2 R} 4 +m \left( \frac 1 {8} -\frac 1
{2\pi} \right) +\frac 1 {128 R} -\frac 1 {24 \pi R} \nn\\
\mbox{res}\ A_3^{(ext)} (-1/2) &=& \frac{2m^3 R^2}{ 3\pi} +m^2 R \left (
 \frac 1 4 -\frac 2 {3\pi} \right) \nn\\
& &-m \left( \frac 1 {8} -\frac 7 {20 \pi} \right)
-\frac 1 {128 R} +\frac{97} {10080 \pi R}. \nn
\eeq
This yields for the residue of the whole zeta function at the exterior region:
\beq
\mbox{Res}\ \zeta^{(ext)} (-1/2) = \frac{1}{63  \pi R} -\frac{m}{15 \pi}
-\frac{m^2R}{3 \pi}  -\frac{m^3R^2}{3\pi}  +\frac{m^4R^3}{6 \pi}. 
\label{polesext}
\eeq
Thus the minimal set of counterterms necessary in order to 
renormalize the theory in the exterior of the bag is identical 
to the one in the interior of the bag. The classical system is 
again described by Eq.~(\ref{n1}).

The opposite sign of the coefficients 
in the divergences (\ref{poles}) and (\ref{polesext})
corresponding to the odd powers
of $R$ can be easily explained by means of differential geometrical
arguments, just  observing that the curvature of the surface of the
bag has opposite sign when looked at from the exterior and from the interior
of the bag. 

Contrary to the scalar case,
the divergences from
the two sides do not anihilate when adding up the two contributions.
In fact, for the zeta function corresponding to the whole space (internal
and external to the bag) we obtain:
\beq
\mbox{Res}\ \zeta (-1/2) =\mbox{Res}\ \zeta^{(int)} (-1/2)+
\mbox{Res}\ \zeta^{(ext)} (-1/2)= -\frac{2m}{15 \pi} 
-\frac{2m^3R^2}{3\pi},
\label{poleswhole}   
\eeq
therefore,  the two free parameters $\sigma$ and $k$ 
remain even if the whole space
is considered. The only exception is the massless field where for this
reason the issue of renormalization is much simpler than in our case.

\section{Conclusions}
\setcounter{equation}{0}
In this paper we have studied in considerable detail a quantum field
theoretical system of a Dirac field with boundary conditions corresponding 
to the MIT bag model. This is the most natural continuation
---in the direction towards approaching truly realistic physical systems---
 of previous work where
only scalar fields were treated \cite{bekl96}. 
We have seen that the consideration of a
fermionic field carries along a number of additional difficulties,
mainly in relation with the 
philosophy of the renormalization process. 
Up to this point, the application of our techniques can be carried out 
essentially in the same way as for the scalar case. 
Starting from the Dirac equation and imposing the boundary conditions we 
have derived an eigenvalue equation in terms of Bessel functions. This basic 
expression has then been  solved, implicitly, in the regions interior 
and exterior
to the bag surface, by using contour integration. This has yielded the
corresponding zeta function in each of the two domains. Extraction of 
the singular 
part of the zeta function has been also done exactly. However, adding
up the contributions of the two parts, not all divergences cancel 
among themselves (as was the case for a scalar field), what theoretically
influences  the playground of the ulterior renormalization process.

In the end, a detailed consideration shows that, after renormalization, two 
dimensionless parameters remain whose values cannot be fixed theoretically, 
but have to be numerically adjusted by direct comparison with the 
physical system
described by the model. In this, we must confess, 
 we are still a bit far from our final
goal. In the sense that, as it stands, our model cannnot be considered
yet to describe a realistic physical situation. This must  be left
to future work, given the complexity of the proposal. In any case,
we should like to point out the rigour and strict systematicity of the 
approach we have used here, and also its relative simplicty, if we compare
it with other methods of similar strength and generality.

As noticeable results of our analysis we would like to mention, that
the Casimir energy may have a clear minimum associated with a stable 
bag radius (see Fig.~1). Comparing this behaviour with the one 
corresponding to the
scalar field, where a maximum occured, this difference
 can clearly be traced back to the
anticommuting nature of the spinor fields, which shows up as a sign in the
definition of the groundstate energy. 

Another interesting observation is that, contrary to the case of parallel
plates, the behaviour of the Casimir energy for large values of $mR$ is not 
exponentially damped. Instead, as is clearly observed from the representations
of the $A_i (s)$ given in the Apps.      A and B, we find a behavior
in inverse powers of the mass. This is directly connected with the 
nonvanishing of the extrinsic curvature at the bag. 

Possible continuation of our approach is in the direction of finite 
temperature and finite densities as considered already for massless
fermions in \cite{francia,hor2,hor3}. 
A natural question to ask  concerns the possible appearance of
 a first order phase transition from
an hadronic bag to a deconfined quark-gluon plasma within our framework.
This is left for future work.

\section*{Acknowledgements}
This investigation has been supported by DGICYT (Spain), project PB93-0035,
by CIRIT (Generalitat de Catalunya), by the Alexander von Humboldt
Foundation, by the German-Spanish program Acciones
Integradas, project HA1995-0171, and by DFG, contract Bo 1112/4-2.

\appendix

\app{Appendix: Explicit representations for the asymptotic contributions
inside the bag}
\setcounter{equation}{0}

The essential formulas for the basic series 
$f(s;a;b)$, eq.~(\ref{fab}), in the calculation are the 
following:
\beq
\sum_{\nu =1/2,3/2,...}^{\infty} \nu^{2n+1} \left(1+\left(\frac{\nu} x
\right) ^2 \right) ^{-s} &=& \frac 1 2 \frac {n! \Gamma (s-n-1)}{\Gamma (s)}
x^{2n+2} \label{continuation1}\\
& &\hspace{-25mm} +(-1)^n 2 \int\limits_0^x d\nu \,\, \frac{\nu^{2n+1} }
{1+e^{2\pi\nu}  } \left( 1-\left( \frac{\nu} x \right)^2 \right) ^{-s} \nn\\
& & \hspace{-25mm} +(-1)^n 2\cos (\pi s)
 \int\limits_x^{\infty} d\nu \,\, \frac{\nu^{2n+1} }
{1+e^{2\pi\nu}  } \left( \left( \frac{\nu} x \right)^2 -1
\right) ^{-s}, \nn
\eeq
\beq
\sum_{\nu =1/2,3/2,...}^{\infty} \nu^{2n} \left(1+\left(\frac{\nu} x
\right) ^2 \right) ^{-s} &=& \frac 1 2 \frac {
\Gamma (n+1/2) \Gamma (s-n-1/2)}{\Gamma (s)}
x^{2n+1} \label{continuation2}\\
& &\hspace{-25mm} -(-1)^n 2\sin (\pi s)
 \int\limits_x^{\infty} d\nu \,\, \frac{\nu^{2n} }
{1+e^{2\pi\nu}  } \left( \left( \frac{\nu} x \right)^2 -1
\right) ^{-s}. \nn
\eeq
Using partial integrations one
can get representations valid for values of $s$ needed 
for the $A_i (s)$.  
One gets, for example, the following expansions around $s=-1/2$:
\beq
\sum_{\nu =1/2,3/2,...}^{\infty} \nu^{3} \left(1+\left(\frac{\nu} x
\right) ^2 \right) ^{-s-3/2} &=& \frac 1 2 \frac {
\Gamma (s-1/2)}{ \Gamma (s+3/2) } x^4   \nn\\
& &\hspace{-35mm} -x^2 \int\limits_0^{\infty}
d\nu \frac{d}{d\nu} \left[ \frac{\nu^2}{ 1+e^{2\pi\nu} }\right]
\ln |\nu^2 -x^2| +{\cal O} (s+1/2) \nn,
\eeq
\beq
\sum_{\nu =1/2,3/2,...}^{\infty} \nu^{2} \left(1+\left(\frac{\nu} x
\right) ^2 \right) ^{-s-3/2} =
-\frac {\pi} 2 x^3 +\pi \frac{x^3}{1+e^{2\pi x} }
+{\cal O} (s+1/2) \nn
\eeq
showing clearly that one can obtain quickly convergent integrals,
respectively expressions for the effective numerical evaluation of the
involved sums.

All the  particular values that are necessary to give the $A_i(s)$, $i=1,2,3$,
explicitly
(in addition to the recurrence (\ref{recur})) are listed in App.     B.
 
The first two leading asymptotic contributions, $A_{-1}$ and $A_0$ have
to be treated in a slightly different way, as has been explained in detail in
\cite{bekl96}. For completeness we give the final results
\beq
A_{-1} (s) &=& \left( \frac{1}{s+1/2} - \ln m^2 \right) \left( -\frac{R^3m^4}{
12\pi} + \frac{m^2R}{24 \pi} + \frac{7}{960 \pi R} \right) \nn \\
&&\hspace{-35mm} + \frac{R^3m^4}{24 \pi} (1-4\ln 2)
- \frac{m^3R^2}{6}  + \frac{m^2R}{24 \pi}
\left[ 2 \ln (2mR) -1 \right]+ \frac{7}{960 \pi R} \left[ 1 +2 \ln (2mR) 
\right] \nn \\
&& -\frac{2}{\pi R} \int_0^\infty \frac{d\nu\, \nu}{1+e^{2\pi \nu}} 
\left( \nu^2 - m^2 R^2 \right)\ln \left| \nu^2 - m^2 R^2 \right|\nn \\
&& -\frac{4m^2R}{\pi} \int_0^\infty \frac{d\nu\, \nu}{1+e^{2\pi \nu}}
\left(
\ln \left| \nu^2 - m^2 R^2 \right| + \frac{\nu}{mR} \ln \left| \frac{mR+ \nu}{
mR -\nu} \right| \right)  \\
&&\hspace{-25mm} + \frac{m^2R}{2 \pi} \ln \left( 1 + e^{-2\pi mR} \right) -\frac{1}{R}
\int_{mR}^\infty\frac{d\nu\, \nu^2}{1+e^{2\pi \nu}}- \frac{m^2R}{\pi}
\int_0^1 dy \, \ln \left( 1 + e^{-2\pi mRy} \right), \nn
\eeq
and 
\beq
A_0 (s) &=& -\left( \frac{1}{s+1/2} - \ln m^2 \right) \left( \frac{1}{48
\pi R} + \frac{m^2 R}{4\pi} \right) \nn \\
&& + \frac{m^3R^2}{6} + \frac{m^2 R}{\pi} \left[ \frac{5}{4} -\frac{1}{2}
\ln 2 - \ln (mR) \right] - \frac{\ln 2}{24\pi R} \nn \\ &&
- \frac{2}{R} \int_{mR}^\infty \frac{d\nu\, \nu^2}{1+e^{2\pi \nu}}-
m^3R^2 \int_0^1 \frac{dx}{1+e^{2\pi mR\sqrt{x}}} \nn \\
&& + \frac{1}{\pi R} \int_0^\infty \frac{d\nu\, \nu}{1+e^{2\pi \nu}}
\ln \left| 1- \left( \frac{\nu}{mR} \right)^2 \right| \nn \\
&& - \frac{m^2R}{2\pi} \int_0^\infty d\nu\, \left( \frac{d}{d\nu} \frac{1}{
1+e^{2\pi \nu}} \right) \int_0^1 \frac{dx}{\sqrt{x}} \, \ln \left| m^2R^2 x-
\nu^2 \right|.
\eeq
This completes the description of our procedures to obtain numerically 
evaluable representations for all the $A_i$'s, needed for the calculation
of the Casimir energy of the spinor inside the bag.

\app{Appendix: Full list of constituent terms $f(a;b)$ 
to be used in addition to the recurrence formula}
\setcounter{equation}{0}

To simplify the expressions, we shall here use $x$ for $mR$. 
In addition to 
the above recurrence, in order to determine all the $A_i(s)$ explicitly one 
needs the following $f(a;b)$'s [we shall use the notation
$f(a;b) = f (-1/2;a;b)$]:
\beq
&&f(0;1/2)= 0,\qquad  f(1;1/2) = -\frac 1 2 x^2 +\frac 1 {24} \nn \\
&&\frac d {ds} \left|_{s=-1/2} \right. f(s;0;1/2)= -\pi x -
2\pi \int_x^{\infty} d\nu \, \frac 1 {1+e^{2\pi \nu}}\nn\\
&&\frac d {ds} \left|_{s=-1/2} \right. f(s;1;1/2) = 
-\frac 1 2 x^2 -2\int_0^{\infty} d\nu 
\frac{\nu} {1+e^{2\pi \nu}} 
\ln \left|1-\left(\frac{\nu}{x}\right)^2 \right| \nn\\
&&f(0;1)= \frac{x}{2(s+1/2)} +x\ln 2 + 2 x^2
\int_x^\infty d\nu \, \frac{d}{d\nu} \left( \frac{1}{\nu
\left( 1+e^{2\pi \nu}\right)}
\right) \, \left[ \left( \frac{\nu}{x} \right)^2-1\right]^{1/2}, \nn \\
&& f(0;3/2)= \frac{\pi x}{2} -\frac{\pi x}{1+e^{2\pi x}}, \nn \\
&&f(1;3/2)=\frac{x^2}{2(s+1/2)} + x^2 \ln x + x^2
\int_0^\infty d\nu\, \left( \frac{d}{d\nu} \frac{1}{
1+e^{2\pi \nu}} \right)\, \ln \left| \nu^2 -x^2 \right|, \nn \\
&&f(1;1)=2x^2 \int_0^x d\nu\, \left( \frac{d}{d\nu} \frac{1}{
1+e^{2\pi \nu}} \right)\, \left[ 1 - \left( \frac{\nu}{x} \right)^2
 \right]^{1/2}, \nn \\
&&f(1;2)=-2x^2 \int_0^x d\nu\, \left( \frac{d}{d\nu} \frac{1}{
1+e^{2\pi \nu}} \right)\, \left| 1 - \left( \frac{\nu}{x} \right)^2
 \right|^{-1/2}, \nn \\
&&f(2;2)=\frac{x^3}{2(s+1/2)}+(\ln 2 -1) x^3 +2x^4 \int_x^\infty d\nu\, 
\left[ \frac{d}{d\nu} \left( \frac{1}{\nu} \frac{d}{d\nu} \frac{\nu}{
1+e^{2\pi \nu}}\right)  \right)\, \left[ \left( \frac{\nu}{x} \right)^2 -1
 \right]^{1/2}, \nn \\
&&f(2;5/2)= \frac{\pi}{4} x^3 - \frac{\pi}{2} x^4 \left. 
\left( \frac{1}{\nu} \frac{d}{d\nu} \frac{\nu}{
1+e^{2\pi \nu}}\right)\right|_{\nu =x}, \nn \\
&&f(3;5/2)=\frac{x^4}{2(s+1/2)}+(\ln x -1/2) x^4 +\frac{x^4}{2}
 \int_0^\infty d\nu\, 
\left[ \frac{d}{d\nu} \left( \frac{1}{\nu} \frac{d}{d\nu} \frac{\nu^2}{
1+e^{2\pi \nu}}\right)  \right]\, \ln \left| \nu^2 - x^2 \right|, \nn \\
&&f(3;3)=  - \frac{2}{3} x^4  \int_0^x d\nu\, 
\left[ \frac{d}{d\nu} \left( \frac{1}{\nu} \frac{d}{d\nu} \frac{\nu^2}{
1+e^{2\pi \nu}}\right)  \right]\,  \left[ 1 - \left( \frac{\nu}{x} \right)^2
 \right]^{-1/2}, \nn \\
&&f(4;3)=\frac{x^5}{2(s+1/2)}+(3\ln 2 -4) \frac{x^5}{3}\nn \\
&& \hspace{3cm} +\frac{2x^6}{3}
 \int_x^\infty d\nu\, 
\left[ \frac{d}{d\nu} \left( \frac{1}{\nu} \frac{d}{d\nu}
 \frac{1}{\nu} \frac{d}{d\nu} \frac{\nu^3}{
1+e^{2\pi \nu}}\right)  \right]\, \left[ \left( \frac{\nu}{x} \right)^2-1
 \right]^{1/2}, \nn \\
&&f(4;7/2)=  \frac{3\pi}{16} x^5 - \frac{\pi}{8} x^6 \left. 
 \left( \frac{1}{\nu} \frac{d}{d\nu}
 \frac{1}{\nu} \frac{d}{d\nu} \frac{\nu^3}{
1+e^{2\pi \nu}}\right)\right|_{\nu =x}, \nn \\
&&f(5;7/2)=\frac{x^6}{2(s+1/2)}+(\ln x -3/4) x^6\nn \\
&& \hspace{3cm} +\frac{x^6}{8}
 \int_0^\infty d\nu\, 
\left[ \frac{d}{d\nu} \left( \frac{1}{\nu} \frac{d}{d\nu}
 \frac{1}{\nu} \frac{d}{d\nu} \frac{\nu^4}{
1+e^{2\pi \nu}}\right)  \right]\, \ln \left| \nu^2 - x^2 \right|, \nn \\
&&f(5;4)= -\frac{2x^6}{15} \int_0^x d\nu\, 
\left[ \frac{d}{d\nu} \left( \frac{1}{\nu} \frac{d}{d\nu}
 \frac{1}{\nu} \frac{d}{d\nu} \frac{\nu^4}{
1+e^{2\pi \nu}}\right)  \right]\, \left[ 1 - \left( \frac{\nu}{x} \right)^2
 \right]^{-1/2}, \nn \\
&&f(6;4)=\frac{x^7}{2(s+1/2)}+(\ln 2 -23/15) x^7 \nn \\
&& \hspace{3cm}   +\frac{2x^8}{15}
 \int_x^\infty d\nu\, 
\left[ \frac{d}{d\nu} \left( \frac{1}{\nu} \frac{d}{d\nu}
 \frac{1}{\nu} \frac{d}{d\nu} \frac{1}{\nu} \frac{d}{d\nu} \frac{\nu^5}{
1+e^{2\pi \nu}}\right)  \right]\, \left[ \left( \frac{\nu}{x} \right)^2-1
 \right]^{1/2}, \nn \\
&&f(6;9/2)=  \frac{5\pi}{32} x^7 - \frac{\pi}{48} x^8 \left. 
 \left( \frac{1}{\nu} \frac{d}{d\nu}\frac{1}{\nu} \frac{d}{d\nu}
 \frac{1}{\nu} \frac{d}{d\nu} \frac{\nu^5}{
1+e^{2\pi \nu}}\right)\right|_{\nu =x}, \nn \\
&&f(7;9/2)=\frac{x^8}{2(s+1/2)}+(\ln x -11/12) x^8 \nn \\ &&
\hspace{3cm}  +\frac{x^8}{48}
 \int_0^\infty d\nu\, 
\left[ \frac{d}{d\nu} \left( \frac{1}{\nu} \frac{d}{d\nu}
 \frac{1}{\nu} \frac{d}{d\nu} \frac{1}{\nu} \frac{d}{d\nu} \frac{\nu^6}{
1+e^{2\pi \nu}}\right)  \right]\, \ln \left| \nu^2 - x^2 \right|.
\eeq
With them, all the $A_i(s)$ are obtained immediately and, what is very
important, always in the most suitable fashion for practical
evaluation (as explained before).


\newpage

\begin{thebibliography}{10}

\bibitem{casimir48}
H.B.G. Casimir, Proc. Koninkl. Ned. Akad. Wetenshap {\bf 51} (1948) 793.

\bibitem{greiner}
G. Plunien, B. M\"uller and W. Greiner, Phys. Rep. {\bf 134} (1986) 87.

\bibitem{vdW1}
E.M. Lifshitz, Soviet Phys. JETP {\bf 2} (1956) 73.

\bibitem{cp1}
H.B.G. Casimir and D. Polder, Phys. Rev. {\bf 73} (1948) 360.

\bibitem{casimir56}
H.B.G. Casimir, 
Physica {\bf 19} (1953) 846.

\bibitem{boyer68}
T.H. Boyer, Phys. Rev. {\bf 174} (1968) 1764.

\bibitem{baldup78}
R. Balian and R. Duplantier, Ann. Phys. {\bf 112} (1978) 165.

\bibitem{mil78}
K.A. Milton, L.L. De Raad Jr. and J. Schwinger, Ann. Phys. {\bf 115} (1978) 388.

\bibitem{eorbz}
E. Elizalde, S.D. Odintsov, A. Romeo, A.A. Bytsenko and S. Zerbini.
{\it Zeta regularization techniques with applications},
World Sci., Singapore (1994).

\bibitem{ee}
E. Elizalde, {\it Ten physical applications of spectral zeta functions},
Springer, Berlin (1995).

\bibitem{cho74}
A. Chodos, R.L. Jaffe. K. Johnson, C.B. Thorn and V. Weisskopf,
Phys. Rev. D {\bf 9} (1974) 3471.

\bibitem{cho74a}
A. Chodos, R.L. Jaffe. K. Johnson and C.B. Thorn,
Phys. Rev. D {\bf 10} (1974) 2599.

\bibitem{bender74}
C.M. Bender and P. Hays, Phys. Rev. D {\bf 14} (1976) 2622.

\bibitem{has78}
P. Hasenfratz and J. Kuti, Phys. Rep. {\bf 40C} (1978) 75.

\bibitem{francia}
M.De Francia, Phys. Rev. D {\bf 50} (1994) 2908.

\bibitem{vep90}
L. Vepstas and A.D. Jackson, Phys. Rep. {\bf 187} (1990) 109;
Nucl. Phys. {\bf A481} (1988) 668.

\bibitem{rho83}
M. Rho, A.S. Goldhaber and G.E. Brown, Phys. Rev. Lett. {\bf 51} (1983)
747.

\bibitem{brow79}
G.E. Brown and M. Rho, Phys. Lett. {\bf B82} (1979) 177.

\bibitem{brow84}
G.E. Brown, A.D. Jackson, M. Rho and V. Vento, Phys. Lett. {\bf B140} 
(1984) 285.

\bibitem{hor1}
M.De Francia, H. Falomir and E.M. Santangelo, Phys. Lett. {\bf B371} (1996) 
285.

\bibitem{hor2}
M.De Francia, H. Falomir and E.M. Santangelo, Phys. Rev. D {\bf 45}
(1992) 2129.

\bibitem{hor3}
M.De Francia, H. Falomir and M. Loewe, Phys. Rev. D {\bf 55} (1997) 2477. 

\bibitem{milton80}
K.A. Milton, Phys. Rev. D {\bf 22} (1980) 1441; {\it ibid} 1444;
Ann. Phys. {\bf 127} (1980) 49.

\bibitem{milton83}
K.A. Milton, Ann. Phys. {\bf 150} (1983) 432.

\bibitem{baacke83}
J. Baacke and Y. Igarashi, Phys. Rev. D {\bf 27} (1983) 460.

\bibitem{andreas}
S.K. Blau, M. Visser and A. Wipf, Nucl. Phys. B {\bf 310} (1988) 163.

\bibitem{bk}
M. Bordag and K. Kirsten, {\it Heat-kernel coefficients of the Laplace operator
in the 3-dimensional ball}, hep-th/9501064.

\bibitem{bek} 
M. Bordag, E. Elizalde and K. Kirsten, J. Math. Phys. {\bf 37} (1996) 895. 

\bibitem{begk}
M. Bordag, E. Elizalde, B. Geyer and K. Kirsten, Commun. Math. Phys. {\bf 179}
(1996) 215.

\bibitem{bdk}
M. Bordag, S. Dowker and K. Kirsten,  Commun. Math. Phys. {\bf 182} (1996) 371.

\bibitem{abdk}
J. Apps, M. Bordag, S. Dowker and K. Kirsten, {\it Spectral invariants for the
Dirac equations on the d-ball with various boundary conditions}, hep-th/9511060.

\bibitem{cg}
G. Cognola and K. Kirsten, Class. Quantum Grav. {\bf 13} (1996) 633.

\bibitem{eli1}  
E. Elizalde, M. Lygren and D.V. Vassilevich,
 J. Math. Phys. {\bf 37} (1996) 3105.

\bibitem{eli2}
E. Elizalde, M. Lygren and D.V. Vassilevich,
Commun. Math. Phys. {\bf 183} (1997) 645.

\bibitem{stuart}
J. Dowker and K. Kirsten, Spinors and forms on generalised cones,
hep-th/9608189.

\bibitem{dowker}
J.S. Dowker and J.S. Apps, Class. Quantum Grav. {\bf 12} (1995) 1363; 
J.S. Dowker, Class. Quantum Grav. {\bf 13} (1996) 1; Phys. Lett. B 
{\bf 366} (1996) 89.

\bibitem{rom}
A. Romeo, Phys. Rev. D {\bf 52} (1995) 7308, {\bf 53} (1996) 3392;
S. Leseduarte and A. Romeo, Europhys. Lett. {\bf 34} (1996) 79-83;
Ann. Phys. (N.Y.) {\bf 250} (1996) 448.

\bibitem{bekl96}
M. Bordag, E. Elizalde, K. Kirsten and S. Leseduarte, Casimir energies 
for massive fields in the bag, hep-th/9608071.

\bibitem{abramo}
M. Abramowitz and I.A. Stegun,
{\it Handbook of Mathematical Functions
  (Natl.~Bur.~Stand.~Appl.~Math.~Ser.55)},
(Washington, D.C.: US GPO),  Dover, New York, reprinted 1972.

  
\end{thebibliography}
\end{document}